\documentclass[12pt,english]{article}

\usepackage[T1]{fontenc}
\usepackage[latin1]{inputenc}
\usepackage{float}
\usepackage{calc}
\usepackage{amsmath}
\usepackage{graphicx}
\usepackage{amssymb}
\usepackage{algorithm}
\usepackage[noend]{algorithmic}
\usepackage{epstopdf, epsf}
\usepackage{natbib}

\makeatletter

\providecommand{\tabularnewline}{\\}

\floatstyle{ruled}
\newfloat{algorithm}{tbp}{loa}
\floatname{algorithm}{Algorithm}

\usepackage{float}

\usepackage{calc}

\makeatletter

\floatstyle{ruled}
\newfloat{algorithm}{tbp}{loa}
\floatname{algorithm}{Algorithm}

\usepackage{times}

\usepackage{epsfig}

\usepackage{amsfonts}

\usepackage{algorithm}

\usepackage[noend]{algorithmic}

  \newtheorem{lem}{Lemma}
  \newtheorem{thm}{Theorem}

\newenvironment{proof}[1][Proof]{\begin{trivlist}
\item[\hskip \labelsep {\bfseries #1}]}{\end{trivlist}}

\usepackage{babel}
\makeatother

\begin{document}

\title{Distinct Counting with a Self-Learning Bitmap}

\author{Aiyou Chen, Jin Cao, Larry Shepp and Tuan Nguyen
\footnote{Aiyou Chen (E-mail: aiyouchen@google.com) is a statistician at Google Inc.,
 Mountain View, CA 94043;
Jin Cao (E-mail: cao@research.bell-labs.com) is a Distinguished Member of
Technical Staff at Bell Labs, Alcatel-Lucent,
Murray Hill, NJ 07974;
Larry Shepp is a Professor, Department of Statistics, Rutgers University,
Piscataway, NJ 08854; and Tuan Nguyen is a Ph.D. candidate, 
Department of Statistics, Rutgers University, Piscataway, NJ 08854. 
The work was done when Aiyou Chen was a Member of Technical Staff at 
Bell Labs, Alcatel-Lucent.     
The authors would like to thank the Editor, AE and two anonymous referees for   
useful reviews and constructive suggestions which have improved the paper 
significantly.
Lawrence Menten at Bell Labs, Alcatel-Lucent proposed an idea of 
hardware implementation for S-bitmap.
}
\vspace{1.6mm}
 \\
Bell Laboratories and Rutgers University
 }

\maketitle
\begin{abstract}
Counting the number of distinct elements (cardinality) in a
dataset is a fundamental problem in database management.  In recent years, due to
many of its modern applications, there has been significant interest
to address the distinct counting problem in a {\em data stream}
setting, where each incoming data can be seen only once and cannot be
stored for long periods of time. 
Many 
probabilistic approaches based on either sampling or sketching
have been proposed in the computer science literature,
that only require limited computing and memory resources.  However,
the performances of these methods are not {\em scale-invariant}, in
the sense that their relative root mean square estimation errors (RRMSE) depend on the unknown cardinalities. This is not
 desirable in many applications where cardinalities can be very
dynamic or inhomogeneous and many cardinalities need to be estimated.  
In this paper, we develop a novel approach, called {\em
self-learning bitmap (S-bitmap)} that is {\it scale-invariant}
for cardinalities in a specified range.  S-bitmap uses a binary vector
whose entries are updated from 0 to 1 by an adaptive sampling process
for inferring the unknown cardinality, where the sampling rates are
reduced sequentially as more and more entries change from 0 to 1. We
prove rigorously that the S-bitmap estimate is not only unbiased but
scale-invariant. We demonstrate
that to achieve a small RRMSE value of
$\epsilon$ or less,
our approach requires
significantly less memory and consumes similar or less operations
than state-of-the-art methods for many common practice cardinality scales.
Both simulation and experimental
studies are reported.
\\

\end{abstract}

\section{Introduction}

\label{sec:intro}
Counting the number of distinct elements (cardinality) in
a dataset is a fundamental problem in database management. In recent years, due
to high rate data collection in many modern applications, there has
been significant interest to address the distinct counting problem in
a {\em data stream} setting where each incoming data can be seen only
once and cannot be stored for long periods of time.  Algorithms to
deal with streaming data are often called {\em online} algorithms.
For example, in modern high speed networks, data traffic in the form
of packets can arrive at the network link in the speed of gigabits
per second, creating a massive data stream. A sequence of packets
between the same pair of source and destination hosts and their
application protocols form a flow, and the number of distinct network
flows is an important monitoring metric for network health (for example, the
early stage of worm attack often results a significant increase in the
number of network flows as infected machines randomly scan others, see
\citet{bu.etc.2006}).
As another example, it is often useful to monitor connectivity
patterns among network hosts and count the number of distinct peers that
each host is communicating with over time \citep{Karasaridis:2007}, in
order to analyze the presence of peer-to-peer networks that are used
for file sharing (e.g. songs, movies).

The challenge of distinct counting in the stream setting is due to the constraint of
 limited memory and computation resources. In this scenario,
the exact solution is infeasible, and a lightweight algorithm, that
derives an {\em approximate} count with low memory and computational cost
but with high accuracy, is desired. In particular, such a solution will be much 
preferred  for counting tasks performed over 
Android-based smart phones (with only limited memory and computing resources), which is
in rapid growth nowadays  \citep{menten.2011}.
Another difficulty is that in many applications, the unknown
cardinalities to be estimated may fall into a wide range, from 1
to $N$, where $N \gg 1$ is a known upper bound.  Hence an algorithm that
can perform uniformly well within the range is preferred.  For
instance,
there can be millions of hosts (e.g. home users) active in a network and the number of
flows each host has may change dramatically from host to host and from
time to time.  
Similarly, a core network may be composed of many links with varying
link speeds, and a traffic snapshot of the network can reveal
variations between links by several orders of magnitude. (A real data
example is given in Section \ref{sec:experiment}.)  It is problematic
if the algorithm for counting number of flows works well (e.g. relative root mean square estimation errors are 
below some threshold) on some links
while not on others due to different scales.

There have been many solutions developed in the computer science
literature to address the distinct counting problem in the stream setting, most notably\cite{FM85}, 
\cite{whang.etc.1990},
\cite{gibbons.2001},
\cite{durand.flajolet.2003},
\cite{bitmap:2006}, 
\cite{flajolet.et.al.07} among others. 
Various asymptotical analyses have been carried out recently, see \cite{kane.etc.2011} and references therein.
The key idea is to obtain a statistical estimate by designing a compact and
easy-to-compute summary statistic (also called sketch in computer
science) from the streaming data.  Some of these methods (e.g.  LogLog
counting by \cite{durand.flajolet.2003} and Hyper-LogLog counting by
\cite{flajolet.et.al.07}) have nice statistical properties such as
asymptotic unbiasedness. However, the performance of these existing
solutions often depends on the unknown cardinalities and cannot
perform uniformly well in the targeted range of cardinalities $[1,N]$.
For example, with limited memory, linear counting proposed by
\cite{whang.etc.1990} works best with small cardinalities while the
LogLog counting method works best with large cardinalities.

Let the performance of a distinct counting method be measured by its
relative root mean square error (RRMSE), where RRMSE is defined by
\begin{eqnarray}  
Re(\hat{n}) & = & \sqrt{\mathbb{E}(n^{-1}\hat{n}-1)^2} \nonumber
\end{eqnarray}
where  $n$ is the
distinct count parameter and $\hat{n}$ is its estimate.  In this article
we develop a novel statistics based distinct counting algorithm,
called {\em S-bitmap}, that is {\it scale-invariant}, in the sense
that RRMSE is invariant to the unknown cardinalities in a wide range
without additional memory and computational costs, i.e. there exists a constant $\epsilon > 0$ such that 
\begin{eqnarray} \label{eq:sinvariance}
Re(\hat{n}) & \equiv &  \epsilon, \text{ for }n=1, \cdots, N.  
\end{eqnarray} 
S-bitmp uses the bitmap, 
i.e., a binary vector, to summarize the data for approximate counting,
where the binary entries are changed from 0 to 1 by an adaptive
sampling process. In the spirit of \cite{morris.1978}, the
sampling rates decrease sequentially as more entries change to 1 with
the optimal rate learned from the current state of the bitmap.  The
cardinality estimate is then obtained by using a non-stationary Markov
chain model derived from S-bitmap.  We use martingale properties to
prove that our S-bitmap estimate is unbiased, and more importantly,
its RRMSE is indeed scale-invariant. Both simulation and experimental
studies are reported. To achieve the same accuracy as state-of-the-art
methods, S-bitmap requires significantly less memory for many common
practice cardinality scales with similar or less computational cost.

The distinct counting problem we consider here is weakly related to
the traditional 'estimating the number of species' problem, see
\cite{bunge.fitzpatrick.1993}, \cite{haas.stokes.1998}, \cite{mao.2006} and references
therein. However, traditional solutions that rely on sample sets of
the population are impractical in the streaming context due to
restrictive memory and computational constraints. While traditional
statistical studies \citep[see][]{BD:01} mostly focus on statistical
inference given a measurement model, a critical new component of the
solution in the online setting, as we study in this paper, is that one
has to design much more compact summary statistics from the data
(equivalent to a model), which can be computed online.

The remaining of the paper goes as follows. Section
\ref{sec:background} further ellaborates the background and reviews
several competing online algorithms from the literature.  Section
\ref{sec:algorithm} and \ref{sec:estimation} describe S-bitmap and
estimation. Section \ref{sec:rule} provides the dimensioning rule for
S-bitmap and analysis. Section \ref{sec:simu} reports simulation
studies including both performance evaluation and comparison with
state-of-the-art algorithms. Experimental studies are reported in
Section \ref{sec:experiment}.  Throughout the paper, $\mathbb{P}$ and
$\mathbb{E}$ denote probability and expectation, respectively, $\ln(x)$ and $\log(x)$ denote the natural logarithm and base-2 logarithm of $x$, and
Table \ref{tb:notation} lists most notations used in the paper.

The S-bitmap algorithm has been successfully implemented in some
Alcatel-Lucent network monitoring products.
A 4-page poster about the basic idea of S-bitmap \citep[see][]{chen.cao.2009}
was presented at the International Conference on Data Engineering in 2009.

\begin{table*}[ht]
\begin{centering}
\begin{tabular}{ll}
\hline
Variable &
Meaning\tabularnewline
\hline
$m$ &
memory requirement in bits\tabularnewline
$n$ &
cardinality to be estimated\tabularnewline
$\hat{n}$ &
S-bitmap estimate of $n$ \tabularnewline
$\mathbb{P}$, $\mathbb{E}$, $var$ &
probability, expectation, variance\tabularnewline
$Re(\hat{n})$ &
$\sqrt{\mathbb{E}(\hat{n}n^{-1}-1)^{2}}$ (relative root mean square error)\tabularnewline
$[0,N]$ &
the range of cardinalities to be estimated\tabularnewline
$C^{-1/2}, \epsilon$ &
(expected, theoretic) relative root mean square error of S-bitmap\tabularnewline
$V$ &
a bitmap vector \tabularnewline
$p_{b}$ &
sequential sampling rate ($1\leq b\leq m$)\tabularnewline
$S_{t}$ &
bucket location in $V$\tabularnewline
$L_{t}$ &
number of 1s in $V$ after the $t$-th distinct item is hashed
into $V$\tabularnewline
$I_{t}$ &
indicator whether the $t$-th distinct item fills in an empty bucket
in $V$\tabularnewline
$\mathcal{L}_{t}$ &
the set of locations of buckets filled with 1s in $V$\tabularnewline
$T_{b}$ &
number of distinct items after $b$ buckets are filled with
1s in $V$\tabularnewline
$t_{b}$ &
expectation of $T_{b}$\tabularnewline
\hline
\end{tabular}
\par\end{centering}

\caption{Some notations used in the paper.}

\label{tb:notation}
\end{table*}

\section{Background}
\label{sec:background}

In this section, we provide some background and review in details a few classes of 
benchmark online distinct counting algorithms from the existing literature that
only require limited memory and computation.  Readers familiar with the area
can simply skip this section.  

\subsection{Overview}
Let ${\cal X}= \{x_1,x_2,\cdots,x_T\}$ be a sequence of items with
possible replicates, where $x_i$ can be numbers, texts, images or
other digital symbols. The problem of distinct counting is to estimate
the number of distinct items from the sequence, denoted as
$n=|\{x_i:1\leq i\leq T\}|$. For example, if $x_i$ is the $i$-th word
in a book, then $n$ is the number of unique words in the book.  It is
obvious that an exact solution can be obtained by listing all distinct
items (e.g. words in the example). However, as we can easily see, this
solution quickly becomes less attractive when $n$ becomes large as it
requires a memory linear in $n$ for storing the list, and an order of
$\log n$ item comparisons for checking the membership of an item in
the list. 

The objective of online algorithms is to process the incoming
data stream in real time where each data can be seen only once, and derive
an approximate count with accuracy guarantees but with a limited storage
and computation budget. 
A typical online algorithm consists of
the following two steps.  First, instead of storing the original data,
one designs a compact sketch such that the essential information about
the unknown quantity (cardinality in this case) is kept. The second
step is an inference step where the unknown quantity is treated as
the parameter of interest, and the sketch is modeled as random variables
(functions) associated with the parameter. In the following, 
we first review a class of
{\it bitmap} algorithms including linear counting by
\cite{whang.etc.1990} and multi-resolution bitmap (mr-bitmap) by \cite{bitmap:2006}, which are closely related to our new
approach. Then we describe another class of Flajolet-Martin type
algorithms.  We also cover other methods briefly such as sampling that do not follow exactly
 the above online sketching framework. 
An excellent review of these and other existing methods can be found in \cite{beyer.etc.2009},
\cite{metwally.etc.2008}, \cite{Gibbons.09}, and in particular,  \cite{metwally.etc.2008} provides extensive simulation comparisons. 
Our new approach will be compared with three state-of-the-art algorithms 
 from the first two classes of methods: mr-bitmap, LogLog counting and Hyper-LogLog counting.

\subsection{Bitmap} \label{subsec:bitmap}
 The bitmap scheme for distinct counting was first proposed in
\cite{astrahan.etc.1987} and then analyzed in details
in \cite{whang.etc.1990}.
To estimate the cardinality of the sequence, the basic idea of {\it
bitmap}, is to first map the $n$ distinct items uniformly randomly  to
$m$ buckets such that replicate items are mapped to the same bucket,
and then estimate the cardinality based on the number of non-empty
buckets.  Here the uniform random mapping is achieved using a
universal hash function \citep[see][]{knuth.1998}, which is essentially a
pseudo uniform random number generator that takes a variable-size
input, called 'key' (i.e. seed), and returning an integer distributed
uniformly in the range of $[1,m]$.\footnote{As an example, by taking the input datum $x$ as  an integer,
the Carter-Wegman hash function is as follows:
$h(x) = ((ax+b) \text{ mod } p) \text{ mod } m$, where $p$ is a large prime, and $a,b$ are two 
arbitrarily chosen integers modulo $p$ with $a\neq 0$. Here $x$ is the key and the output is an integer in $\{1, \cdots, m\}$ if we replace 0 with $m$.}
To be convenient, let $h: \mathcal{X}\rightarrow \{1,\cdots,m\}$ be a
universal hash function, where it takes a {\em key} $x\in\mathcal{X}$
and map to a hash value $h(x)$.  For theoretical analysis, we assume
that the hash function distributes the items randomly, e.g. for any
$x,y\in \mathcal{X}$ with $x\neq y$, $h(x)$ and $h(y)$ can be treated
as two independent uniform random numbers. A bitmap of length $m$ is
simply a binary vector, say $V=(V[1],\ldots,V[k], \ldots, V[m])$ where
each element $V[k]\in \{0,1\}$.

The basic bitmap algorithm for online distinct counting is as
follows. First, initialize $V[k]=0$ for $k=1,\cdots,m$. Then for each
incoming data $x \in \mathcal{X}$, compute its hash value $k=h(x)$
and update the corresponding entry in the bitmap $V[k]$ by setting
$V[k]=1$. For convenience, this is summarized in Algorithm \ref{alg:bitmap}.   Notice that the bitmap algorithm requires a storage of $m$
bits attributed to the bitmap and requires no additional storage for
the data.  It is easy to show that each entry in the bitamp $V[k]$ is
$Bernoulli(1-\left(1-m^{-1}\right)^n)$, and hence the distribution of
$|V|=\sum_{k=1}^mV[k]$ only depends on $n$.  Various estimates of $n$
have been developed based on $|V|$, for example, linear counting as
mentioned above uses the estimator $m\ln(m(m-|V|)^{-1})$. The name
'linear counting' comes from the fact that its memory requirement is
almost linear in $n$ in order to obtain good estimation.

\begin{algorithm}[tttt]
\caption{Basic bitmap}
\begin{algorithmic}[1]
\begin{small}
\item[Input: a stream of items $x$]
\item[\hspace{.25in} $V$ (a bitmap vector of zeros with size $m$)]
\item[Output: $|V|$ (number of entries with 1s in $V$)]
\item[Configuration: $m$]
     \FOR {$x \in\mathcal{X}$}
        \STATE  compute its hash value $k = h(x)$
        \IF{ $V[k] = 0$}
        \STATE update $V[k] = 1$
        \ENDIF
    \ENDFOR
\STATE Return $|V| = \sum_{k=1}^m V[k]$.
\end{small}
\end{algorithmic}
\label{alg:bitmap}
\end{algorithm}

Typically, $N$ is much larger than the required memory $m$ (in bits),
thus mapping from $\{0,\cdots,m\}$ to $\{1,\cdots,N\}$ cannot be
one-to-one, i.e. perfect estimation, but one-to-multiple.  A bitmap of
size $m$ can only be used to estimate cardinalities less than $m\log
m$ with certain accuracy.
In order to make it scalable to a larger cardinality scale, a
few improved methods based on bitmap have been developed 
\citep[see][]{bitmap:2006}.
One method, called {\it virtual bitmap}, is to apply the bitmap scheme
on a subset of items that is obtained by sampling original items with
a given rate $r$. Then an
estimate of $n$ can be obtained by estimating the cardinality of the sampled
subset.
But it is impossible for virtual bitmap with a single $r$ to
estimate a wide range of
cardinalities accurately.  \cite{bitmap:2006} proposed
a multiresolution bitmap ({\it mr-bitmap}) to improve virtual
bitmap. The basic idea of mr-bitmap is to
make use of multiple virtual
bitmaps, each with a different sampling rate, and embeds
them into one bitmap in a memory-efficient way. To be precise,
it first partitions the original bitmap into $K$ blocks (equivalent to
$K$ virtual bitmaps), and then
associates buckets in the $k$-th block with a sampling rate $r_k$ for
screening distinct items. It may be worth pointing out that mr-bitmap determines $K$ and the
sampling rates with a quasi-optimal strategy and it is still an open question how to 
optimize them, which we leave for future study. Though there is no
rigorous analysis in \cite{bitmap:2006}, mr-bitmap is
not scale-invariant as suggested by simulations in Section \ref{sec:simu}.

\subsection{Flajolet-Martin type algorithms}
The approach of \cite{FM85} (FM) has pioneered a different class of algorithms.
The basic idea of FM  is to first map each item $x$ to a geometric random number $g$, 
and then record the maximum value of the geometric random numbers $\max(g)$,
which can be updated sequentially. In the implementation of FM, upon the arrival of an item $x$, the corresponding $g$ is the location of the left-most 1 in the binary vector $h(x)$ (each entry of the binary vector follows $Bernoulli(1/2)$), where $h$ is a universal hash function mentioned earlier. Therefore $\mathbb{P}(g=k) = 2^{-k}$.
Naturally by hashing, replicate items are mapped to the same geometric random number.
The maximum order statistic $\max(g)$ is the summary statistic for FM, 
also called the FM sketch in the literature. 
Note that the distribution of $\max(g)$ 
is completely determined by the number of distinct items. By randomly partitioning items into $m$ groups, the FM approach
obtains $m$ maximum random numbers, one for each group, which are independent and identically distributed, and then estimates
the distinct count by a moment method. 
Since FM makes use of the binary value of $h(x)$, which requires at most $\log(N)$ bits of memory
where
$N$ is the upper bound of distinct counts (taking as power of 2), it is also called {\em log-counting}.
Various extensions of the FM approach have been explored in the literature based on the $k$-th maximum order statistic, where $k=1$ corresponds to FM  \citep[see][]{giroire.2005, beyer.etc.2009}.

Flajolet and his collaborators have recently proposed two innovative methods, 
called LogLog counting and Hyper-LogLog as mentioned above, published in 2003 and 2007, subsequently.
Both methods use the technique of
recording  the binary value of $g$ directly, which requires at most $\log(\log N)$ bits (taking $N$ such that $\log(\log N)$ is integer), and therefore are also called loglog-counting. This 
provides a more compact summary statistic than FM. 
Hyper-LogLog is built on a more efficient estimator  than LogLog, see \cite{flajolet.et.al.07} for the exact formulas of the estimators. 

Simulations suggest that although Hyper-LogLog may have a bounded RRMSE for
cardinalities in a given range, its RRMSE fluctuates as
cardinalities change and thus it is {\em not} scale-invariant.

\subsection{Distinct sampling}
The paper of \cite{flajolet.1990} proposed a novel sampling algorithm, called {\em Wegman's adaptive sampling}, which
collects a random sample of the distinct elements (binary values) of size no more than a pre-specified number. Upon arrival of a new distinct element, if the sample size of the existing collection is more than a threshold, the algorithm will remove some of the collected sample and the new element will be inserted with a sampling rate $2^{-k}$, where $k $ starts from 0 and grows adaptively according to available memory. The {\em distinct sampling} of \cite{gibbons.2001} uses the same idea to collect a random sample of distinct elements. 
These sampling algorithms are essentially different from the above two classes of algorithms based on one-scan sketches, and 
are computationally less attractive as they require scanning all existing collection periodically.  They belong to the log-counting family with memory cost in the order of $\epsilon^{-2}\log(N)$ where $\epsilon$ is an asymptotic RRMSE, but their asymptotic memory efficiency  is somewhat worse than the original FM method, see \cite{flajolet.et.al.07} for an asymptotic comparison.
\cite{flajolet.1990} has shown that with a finite population, the RRMSE of Wegman's adaptive sampling exhibits periodic fluctuations, depending on unknown cardinalities, and thus it is not scale invariant as defined by 
\eqref{eq:sinvariance}.
Our new approach makes use of the general idea of adaptive sampling, but is quite different from these sampling algorithms, as ours does not require collecting a sample set of distinct values, and furthermore is scale invariant as shown later.

\section{Self-learning Bitmap}

\label{sec:algorithm}

As we have explained in Section \ref{subsec:bitmap}, the basic bitmap (see Algorithm \ref{alg:bitmap}), as well as virtual bitmap, provides a memory-efficient data summary but they cannot be used to estimate cardinalities accurately in a wide range. In this section, we describe a new approach for online distinct counting by building a {\em self-learning} bitmap (S-bitmap for abbreviation), which not only is memory-efficient, but  provides a scale-invariant estimator with high accuracy.

The basic idea of S-bitmap is to build an adaptive sampling process
into a bitmap as our summary statistic, where
the sampling rates  decrease
sequentially as more and more new distinct
items arrive.  The motivation for decreasing
sampling rates is easy to perceive - if one draws Bernoulli sample
with rate $p$ from a
population with unknown size $n$ and obtains a Binomial count, say $X\sim Binomial(n,p)$, then
the maximum likelihood estimate $p^{-1}X$ for $n$ has relative mean
square error
$\mathbb{E}(n^{-1}p^{-1}X-1)^2=(1-p)/(np)$. So, to achieve a constant
relative error, one needs to use a smaller sampling rate $p$ on a larger
population with size $n$. The sampling idea is similar to
``adaptive sampling'' of \cite{morris.1978} which was proposed for
counting a large number of items with {\it no item-duplication} using a small
memory space. However, since the main issue of distinct counting is
item-duplication,  Morris' approach does not apply
here. 

\begin{figure*}[ht]
\centering
\includegraphics[width=4.5in,height=3.5in]{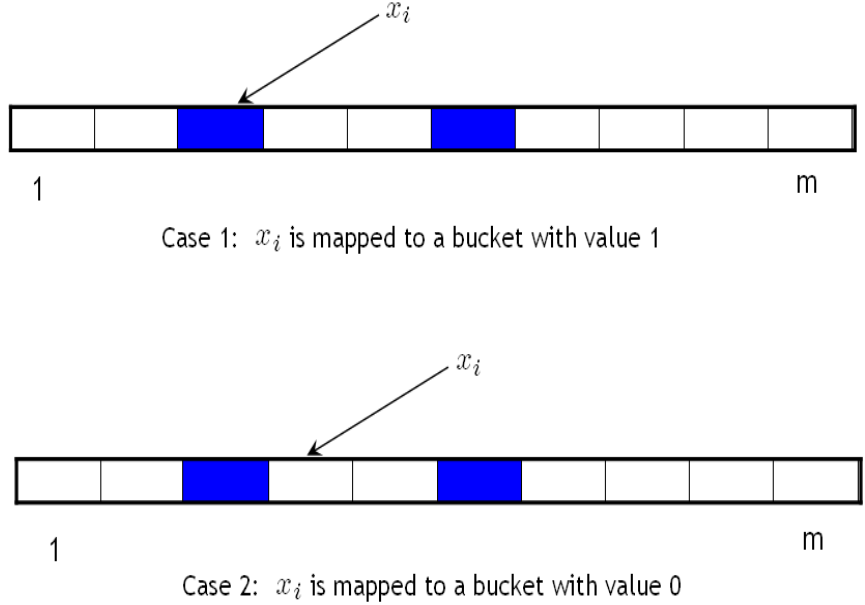}
\caption{Update of the bitmap vector: in case 1, just skip to the
  next item, and in case 2, with probability $p_L$ where $L$ is the
  number of 1s in $V$ so far, the bucket value
  is changed from 0 to 1.}\label{fig:sbitmap.fig}
\end{figure*}

Now we describe S-bitmap and show how it
deals with  the item-duplication issue effectively.
The basic algorithm for extracting the S-bitmap summary statistic is
as follows. Let $1\geq p_1\geq p_2\geq \cdots \geq p_m>0$ be specified
sampling rates.
A bitmap vector $V\in\{0,1\}^{m}$ with length $m$ is initialized
with 0 and a counter $L$ is initialized by 0 for the number of buckets
filled with 1s. Upon the arrival of a new item $x$ (treated as a
string or binary vector), it is
mapped, by a
universal hash function using $x$ as the key, to say
$k\in\{1,\cdots,m\}$.
If $V[k]=1$, then skip to the next item; Otherwise, with probability
$p_{L}$, $V[k]$ is changed from 0 to 1, in which case $L$ is increased by 1. (See
Figure \ref{fig:sbitmap.fig} for an illustration.)
Note that the sampling is also realized with a universal hash function
using $x$ as keys. Here, $L\in\{0,1,\cdots,m\}$ indicates how many
1-bits by the end of the stream update.
Obviously, the bigger $L$ is, the larger the cardinality is expected
to be. We show in Section \ref{sec:estimation} how to use $L$ to
characterize the distinct count.

If $m=2^c$ for some integer $c$, then S-bitmap can be implemented efficiently
as follows. Let $d$ be an integer. For
each item $x$, it is mapped by a universal hash function using $x$ as
the key to a binary vector with length $c+d$. Let $j$ and $u$ be two integers
that correspond to the binary representations with the first $c$
bits and last $d$ bits, respectively. 
Then $j$ is the bucket location
in the bitmap that the item is hashed into, and $u$ is used for
sampling.  It is easy to see that $j$ and $u$ are independent.  If the bucket is empty,
i.e. $V[j]=0$,  then check whether $u2^{-d}< p_{L+1}$ and if true,
update $V[j]=1$. If the bucket is not empty, then just skip to next item.
This is summarized in Algorithm
\ref{alg:sbitmap1}, where the choice of $(p_1,\cdots,p_m)$ is
described in Section \ref{sec:rule}. Here we follow the setting of
the LogLog counting paper by \cite{durand.flajolet.2003} and take
$\mathcal{X}=\{0,1\}^{c+d}$. There is a chance of collision for hash
functions.  Typically $d=30$, which is small relative to $m$, is sufficient for
$N$ in the order of millions.

\begin{algorithm}[tttt]
\caption{S-bitmap (SKETCHING UPDATE)}
\begin{algorithmic}[1]
\begin{small}
\item[Input: a stream of items $x$ (hashed binary vector with size
  $c+d$)]
\item[\hspace{.25in} $V$ (a bitmap vector of zeros with size $m=2^c$)]
\item[Output: $B$ (number of buckets with 1s in $V$)]
\item[Configuration: $m$]
\STATE Initialize $L=0$
    \FOR {$x=b_{1}\cdots b_{c+d}\in\mathcal{X}$}
        \STATE set $j:=[b_1\cdots b_c]_2$ (integer value of first $c$ bits
        in base 2)
          \IF {$V[j]=0$}
            \STATE $u=[b_{c+1}\cdots b_{c+d}]_2$ 
            \STATE \# sampling \#
            \IF {$u2^{-d}< p_{L+1}$}
               \STATE $V[j]=1$
               \STATE $L=L+1$
            \ENDIF
          \ENDIF
    \ENDFOR
\STATE Return $B=L$.
\end{small}
\end{algorithmic}
\label{alg:sbitmap1}
\end{algorithm}

Since the sequential sampling rates $p_{L}$ only depend on $L$ which
allows us to learn the number of distinct items already passed, the
algorithm is called Self-learning bitmap (S-bitmap). \footnote{Statistically, the self 
learning process can also be called adaptive sampling. We notice that  \cite{bitmap:2006} have used
'adaptive bitmap' to stand for a virtual bitmap where the sampling rate is chosen adaptively based on 
another rough estimate, and that \cite{flajolet.1990} has used 'adaptive sampling' for subset sampling. 
To avoid potential confusion with these,
we use the name 'self learning bitmap' instead of 'adaptive sampling bitmap'.}
We note that the decreasing property of the sampling rates, beyond the
above heuristic optimality, is also
sufficient and necessary for filtering out all duplicated items. To see the
sufficiency, just note
if an item is not
sampled in its first appearance, then the
$d$-bits number associated with
it (say $u$, in line 5 of Algorithm \ref{alg:sbitmap1}) is larger than its current sampling rate, say $p_L$. Thus its later
replicates, still mapped to $u$,  will
not be sampled either due to the monotone property. Mathematically,  if the item is mapped to $u$ with $u2^{-d} > p_L$, then $u2^{-d} > p_{L+1}$ since $p_{L+1} \leq p_L$.
On the other hand, if $p_{L+1} > p_L$, 
then in line 7 of Algorithm \ref{alg:sbitmap1}, $\mathbb{P}(p_L < u2^{-d} <p_{L+1}) >0 $,  that is,
there is a positive probability that the item mapped to $u$, in its first appearance, is not sampled at $L$, but its later replicate is sampled at $L+1$, 
which
establishes the necessity.
The argument of sufficiency here will be used to derive S-bitmap's
Markov property in Section \ref{subsec:markov} which leads to the
S-bitmap estimate of the distinct count using $L$.

It is interesting to see that unlike mr-bitmap, the sampling rates for
S-bitmap are not associated with the bucket locations, but only depend
on the arrival of new distinct items, through increases of $L$. In
addition, we use the memory more efficiently since we can adaptively
change the sampling rates to fill in more buckets, while mr-bitmap may
leave some virtual bitmaps unused or some completely filled, which
leads to some waste of memory.

We further note that in the S-bitmap update process, only one hash is needed
for each incoming item. For bucket update, only
if the mapped bucket is empty, the last $d$-bits of the hashed value
is used to determine whether the bucket should be filled with
1 or not. Note that the sampling rate changes only when an empty bucket
is filled with 1. For example, if $K$ buckets become filled by the
end of the stream, the sample rates only need to be updated $K$ times.
Therefore, the computational cost of S-bitmap is very low, and is
similar to or lower than that of benchmark algorithms such as
mr-bitmap, LogLog and Hyper-LogLog (in fact,
Hyper-LogLog uses the same summary statistic as LogLog and thus their
computational costs are the same).

\section{Estimation}\label{sec:estimation}

In this section, we first derive a Markov chain model for the above $L$
sequence and then obtain the S-bitmap estimator.

\subsection{A non-stationary Markov chain model}\label{subsec:markov}

From the S-bitmap update process, it is clear that the $n$
distinct items are randomly mapped into the $m$ buckets, but not
all corresponding buckets have values 1.
From the above sufficiency argument, due to decreasing sampling rates,
the bitmap filters out replicate
items automatically and its update only depends on the first arrival
of each distinct item, i.e. new item. Without loss of generality, let the $n$ distinct
items be hashed into locations $S_{1},S_{2},\cdots,S_{n}$ with $1\leq
S_i \leq m$, indexed
by the sequence of their first arrivals. Obviously, the $S_{i}$
are i.i.d.. Let $I_{t}$ be the indicator of whether or not the $t$-th distinct item fills
an empty bucket with 1. In other words, $I_{t}=1$ if and only if
the $t$-th distinct item is hashed into an empty bucket (i.e. with
value 0) and further fills
it with 1. Given the first $t-1$ distinct items, let
$\mathcal{L}(t-1)=\{ S_{j}:I_{j}=1,1\leq j\leq t-1\}$
be the buckets that are filled with 1, and $L_{t-1}=|\mathcal{L}(t-1)|$
be the number of buckets filled with 1. Then $L_{t}=L_{t-1}+I_{t}$.  Upon the arrival of the $t$-th
distinct item that is hashed to bucket location $S_t$, if
$S_{t}$ does not belong to $\mathcal{L}(t-1)$, i.e, the bucket is empty,
then by the design of S-bitmap, 
$I_{t}$ is independent of $S_t$.
To be precise, as defined in line 3 and 5 of Algorithm \ref{alg:sbitmap1}, $j$ and $u$ associated with $x$ are independent, one determining the location $S_t$ and the other determining sampling $I_t$.
Obviously, according to line 7 of Algorithm \ref{alg:sbitmap1},  the conditional
probability that the $t$-th distinct item fills the $S_{t}$-th bucket with 1 is
$p_{L_{t-1}+1}$, otherwise is 0,  that is,
\begin{eqnarray*}
\mathbb{P}(I_{t}=1|S_{t}\notin\mathcal{L}(t-1),L_{t-1}) & = &
p_{L_{t-1}+1}\end{eqnarray*}
and
 \begin{eqnarray*}\label{eq:probI}
\mathbb{P}(I_{t}=1|S_{t}\in\mathcal{L}(t-1),L_{t-1}) & = & 0.
\end{eqnarray*}
 The final output from the update algorithm is denoted by $B$, i.e.
\begin{eqnarray*}
B & \equiv & L_{n}  = \sum_{t=1}^n I_t,\end{eqnarray*}
 where $n$ is the parameter to be estimated.

Since $S_{t}$ and $\mathcal{L}(t-1)$
are independent, we have \begin{eqnarray*}
 &  & \mathbb{P}(I_{t}=1|L_{t-1})\\
 &  & =\mathbb{P}(I_{t}=1|S_{t}\notin\mathcal{L}(t-1),L_{t-1})\mathbb{P}(S_{t}\notin\mathcal{L}(t-1)|L_{t-1})\\
 &  & =p_{L_{t-1}+1}\cdot(1-\frac{L_{t-1}}{m}).\end{eqnarray*}
This leads to the Markov chain
property of $L_t$ as summarized in the
theorem below.

\begin{thm}\label{thm:mc}
 Let  $q_{k} =  (1-m^{-1}(k-1))p_{k}$ for $k=1,\cdots,m$.
If the monotonicity condition holds, i.e. $p_1\geq p_2\geq \cdots$,  then
$\{ L_{t}:t=1,\cdots,n\}$
follows a non-stationary Markov chain model: \begin{eqnarray*}
L_{t} & = & L_{t-1}+1,\text{ with probability }q_{L_{t-1}+1}\\
 & = & L_{t-1},\text{ with probability }1-q_{L_{t-1}+1}.\end{eqnarray*}
\end{thm}

\subsection{Estimation}

Let $T_{k}$ be the index for the distinct item that fills an empty
bucket with 1 such that there are $k$ buckets filled with 1 by that
time. That is, $\{ T_{k}=t\}$ is equivalent to \{$L_{t-1}=k-1$ and
$I_{t}=1$\}.
Now given the output $B$ from the update algorithm, obviously
$T_{B}\leq n<T_{B+1}$. A natural estimate of $n$ is
\begin{eqnarray}\label{eq:nhat}
\hat{n} &=& t_{B},
\end{eqnarray}
where
$t_{b}=\mathbb{E}T_{b}$, $b=1,2,\cdots$.

Let $T_{0}\equiv0$ and $t_{0}=0$
for convenience. The following properties hold for $T_b$ and $t_b$.

\begin{lem} \label{lem:geomlem}
Under the monotonicity condition of $\{p_k\}$, $T_{k}-T_{k-1}$, for
$1\leq k\leq m$  are distributed independently with geometric
distributions, and
for $1\leq t\leq m$, \begin{eqnarray*}
\mathbb{P}(T_{k}-T_{k-1}=t) & = & (1-q_{k})^{t-1}q_{k}.\end{eqnarray*}
 The expectation and variance of $T_b$, $1\leq b\leq m$ can be expressed
as \begin{eqnarray*}
t_{b} & = & \sum_{k=1}^{b}q_{k}^{-1}.\end{eqnarray*}
 and \begin{eqnarray*}
var(T_{b}) & = & \sum_{k=1}^{b}(1-q_{k})q_{k}^{-2}.\end{eqnarray*}
 \end{lem}

The proof of Lemma \ref{lem:geomlem} follows from the standard Markov chain
theory and is provided in the appendix for completeness. Below we
analyze how to choose the sequential sampling rates $\{ p_{1},\cdots,p_{m}\}$
such that $Re(\hat{n})$ is stabilized for arbitrary
$n\in\{1,\cdots,N\}$.

\section{Dimensioning rule and analysis}

\label{sec:rule}

In this section, we first describe the dimensioning rule for choosing the
sampling rates $\{p_k\}$.
Notice that $T_{b}$ is an unbiased estimate of
$t_{b}=\mathbb{E}T_{b}$ if $T_b$ is observable but $t_b$ is
unknown, where $t_{1}<t_{2}<\cdots<t_{m}$. Again formally denote
$Re(T_{b})=\sqrt{\mathbb{E}(T_{b}t_{b}^{-1}-1)^{2}}$ as the relative
error. In order to make the RRMSE
 of S-bitmap invariant to the unknown cardinality $n$, our idea is to choose
the sampling rates $\{p_k\}$ such that $Re(T_{b})$ is invariant for $1\leq
b\leq m$, since
$n$ must fall in between some two consecutive $T_{b}$s.
We then prove that although
$T_{b}$ are unobservable, choosing parameters that stabilizes
$Re(T_{b})$ is sufficient for stabilizing the RRMSE of S-bitmap
for all $n\in \{1,\cdots,N\}$.


\subsection{Dimensioning rule}\label{subsec:dimrule}

To stabilize $Re(T_{b})$, we need some constant $C$ such that for
$b=1,\cdots,m$,
\begin{eqnarray}
Re(T_{b}) & \equiv & C^{-1/2}.\label{eq:stable}\end{eqnarray}
This leads to the dimensioning rule for S-bitmap
as summarized by the following theorem, where $C$ is determined later as a
function of $N$ and $m$.

\begin{thm} Let $\{T_k-T_{k-1}:1\leq k\leq m\}$ follow independent
 Geometric distributions as in Lemma \ref{lem:geomlem}. Let
 $r=1-2(C+1)^{-1}$. If
\begin{eqnarray*}
p_{k}=\frac{m}{m+1-k}(1+C^{-1})r^{k},
\end{eqnarray*}
then
we have for $k=1,\cdots,m,$\begin{eqnarray}\label{eq:stableT}
\frac{\sqrt{var(T_{k})}}{\mathbb{E}T_{k}} & \equiv & C^{-1/2}.\end{eqnarray}
 That is, the relative errors $Re(T_{b})$ do not depend on $b$. \label{thm:design} \end{thm}

\begin{proof} Note that \eqref{eq:stableT} is equivalent to \begin{eqnarray*}
\frac{var(T_{b+1})}{t_{b+1}^{2}} & = & \frac{var(T_{b})}{t_{b}^{2}}.\end{eqnarray*}
 By Lemma \ref{lem:geomlem}, this is equivalent to \begin{eqnarray*}
\frac{var(T_{b})+(1-q_{b+1})q_{b+1}^{-2}}{(t_{b}+q_{b+1}^{-1})^{2}} & = & \frac{var(T_{b})}{t_{b}^{2}}.\end{eqnarray*}
 Since $var(T_{b})=C^{-1}t_{b}^{2}$, then
\begin{eqnarray}
q_{b+1}^{-1} & = & \frac{C}{C-1}+\frac{2t_{b}}{C-1}.\label{eq:seqqb}\end{eqnarray}
 Since $t_{b+1}=t_{b}+q_{b+1}^{-1}$, we have \begin{eqnarray*}
t_{b+1} & = & \frac{C+1}{C-1}t_{b}+\frac{C}{C-1}.\end{eqnarray*}
By deduction, \begin{eqnarray*}
t_{b+1} & = & \left(\frac{C+1}{C-1}\right)^{b}\left(t_{1}+2^{-1}C\right)-\frac{C}{2}.\end{eqnarray*}
 Since $var(T_{1})=(1-q_{1})q_{1}^{-1}=C^{-1}t_{1}^{2}$ and $t_{1}=q_{1}^{-1}$,
we have $t_{1}=C(C-1)^{-1}$. Hence with some calculus, we have, for
$r=1-2(C+1)^{-1}$, \begin{eqnarray*}
t_{b} & = & \frac{C}{2}(r^{-b}-1)\end{eqnarray*}
 \begin{eqnarray*}
q_{b} & = & (1+C^{-1})r^{b}.\end{eqnarray*}
 Since $q_{b}=(1-\frac{b-1}{m})p_{b}$, the sequential sampling rate
$p_{b}$, for $b=1,\cdots,m$, can be expressed as \begin{eqnarray*}
p_{b} & = & \frac{m}{m+1-b}(1+C^{-1})r^{b}.\end{eqnarray*}
 The conclusion follows as the steps can be reversed.
\end{proof}

It is easy to
check that the monotonicity property holds strictly for $\{p_k: 1\leq k\leq
m-2^{-1}C\}$, thus satisfying the condition of Lemma
\ref{lem:geomlem}. For $k>m-2^{-1}C$, the monotonicity does not
hold. So it is natural to expect that the upper bound $N$ is
achieved when $m-2^{-1}C$ buckets (suppose $C$ is even) in the bitmap turn into 1, i.e.
$t_{m-2^{-1}C}=N$, or,
 \begin{eqnarray}
N & = &
\frac{C}{2}\left(r^{-(m-2^{-1}C)}-1\right)
.\label{eq:Nmax}\end{eqnarray}
Since $r=1-2(C+1)^{-1}$, we obtain
\begin{eqnarray}
m & = & \frac{C}{2}+\frac{\ln(1+2NC^{-1})}{\ln(1+2(C-1)^{-1})}.\label{eq:m}\end{eqnarray}
Now, given the maximum possible cardinality $N$ and bitmap size $m$, $C$ can be solved uniquely from this equation.

For example, if $N=10^{6}$ and $m=30,000$ bits, then from \eqref{eq:m}
we can solve $C\approx 0.01^{-2}$.
That is, if the sampling rates $\{p_k\}$ in Theorem
\ref{thm:design} are designed using such $(m,N)$, then $Re(\hat{n})$
 can be expected to be approximately 1\% for all
 $n\in\{1,\cdots,10^{6}\}$. In other words, to achieve errors no more
 than 1\% for all possible cardinalities from 1 to $N$, we need only
 about 30 kilobits memory for S-bitmap. 
 
 Since $\ln(1+x) \approx x(1-\frac{1}{2}x)$ for $x$ close to 0,
   \eqref{eq:m} also implies that to achieve a small RRMSE $\epsilon$, which is equal to $(C-1)^{-1/2}$ according to Theorem \ref{thm:accuracy} below, the memory requirement can be approximated as follows:
\begin{eqnarray}
 m & \approx &  \frac{1}{2}\epsilon^{-2}(1+\ln (1+2N\epsilon^2)). \nonumber
 \end{eqnarray}
 Therefore, asymptotically, the memory efficiency of S-bitmap is much better than log-counting algorithms which requires a memory in the order of $\epsilon^{-2}\log N$.   Furthermore, assuming $N \epsilon^{-2} \gg 1$, if $\epsilon < \sqrt{(\log N)^\eta/(2eN)}$ where $\eta\approx 3.1206$, S-bitmap is better than Hyper-LogLog counting which requires memory approximately $1.04^2 \epsilon^{-2} \log(\log N)$ \citep[see][]{flajolet.et.al.07} in order to achieve an asymptotic RRMSE 
 $\epsilon$, otherwise is worse than Hyper-LogLog.

{\bf Remark.}
In implementation, we set
$p_b\equiv p_{m-2^{-1}C}$ for $m-2^{-1}C
\leq b\leq m$ so that the sampling rates satisfy the monotone property
which is necessary by Lemma \ref{lem:geomlem}.
Since the focus is on cardinalities in the range from 1
to $N$ as pre-specified, which corresponds to $B\leq m-2^{-1}C$ as
discussed in the above, we simply truncate the output $L_n$ by $m-2^{-1}C$
if it is larger than this value which becomes possible when $n$ is
close to $N$, that is,
\begin{eqnarray}\label{eq:truncB}
B &=& \min(L_n, m-2^{-1}C).
\end{eqnarray}

\subsection{Analysis}

Here we prove that the S-bitmap estimate is unbiased and its relative
estimation error is indeed ``{\it scale-invariant}`` as we had
expected if we ignore the truncation effect in (\ref{eq:truncB}) for
simplicity.


\begin{thm} \label{thm:accuracy}
Let $B=L_n$, where $L_n$ is the number of 1-bits in the S-bitmap, as
defined in Theorem \ref{thm:mc} for $1\leq n\leq N$. Under the
dimensioning rule of Theorem \ref{thm:design}, for the
S-bitmap estimator $\hat{n}=t_B$ as defined in \eqref{eq:nhat}, we have
\begin{eqnarray*}
\mathbb{E}\hat{n} & = & n\end{eqnarray*}
 \begin{eqnarray*}
\text{RRMSE}(\hat{n}) & = & (C-1)^{-1/2}.\end{eqnarray*}

\end{thm}

\begin{proof} Let for $a>1$ \begin{eqnarray*}
Y_{n} & = & \prod_{j=0}^{L_{n}}(1+(a-1)q_{j}^{-1}).\end{eqnarray*}
By Theorem \ref{thm:mc}, $L_{n+1}=i+Bernoulli(q_{i+1})$ if $L_{n}=i$. Thus \begin{eqnarray*}
 &  & \mathbb{E}(Y_{n+1}|Y_{0},Y_{1},\cdots,Y_{n})\\
 & = & \mathbb{E}(Y_{n}I(L_{n+1}=i)+Y_{n}(1+(a-1)q_{L_{n+1}}^{-1})I(L_{n+1}=i+1)|L_{n})\\
 & = & Y_{n}\{1-q_{i+1}+q_{i+1}(1+(a-1)q_{i+1}^{-1})\}\\
 & = & Y_{n}a\end{eqnarray*}
 if $L_{n}=i$. Therefore $\{ a^{-n}Y_{n}:n=0,1,\cdots\}$ is a martingale.

Note that $q_{i}=(1+C^{-1})r^{i}$, $i\geq0$, where $r=1-2(C+1)^{-1}$.
Since $L_{0}=0$, $\mathbb{E}Y_{0}=1+(a-1)q_{0}^{-1}$
and since $a^{-n}\mathbb{E}Y_{n}=\mathbb{E}Y_{0}$, we have \begin{eqnarray*}
\mathbb{E}Y_{n} & = & a^{n}(1+(a-1)q_{0}^{-1})\end{eqnarray*}
 that is, \begin{eqnarray*}
a^{n}(1+(a-1)q_{0}^{-1}) & = & \mathbb{E}\prod_{j=0}^{L_{n}}(1+(a-1)q_{j}^{-1}).\end{eqnarray*}
 Recall that $t_{b}=\sum_{j=1}^{b}q_{j}^{-1}$ and
 $\sum_{j=1}^{b}q_{j}^{-2}(1-q_{j})=C^{-1}(\sum_{j=1}^{b}q_{j}^{-1})^{2}$.
Taking first derivative at $a=1_+$, we have (since $B=L_{n}$) \begin{eqnarray*}
n+q_{0}^{-1} & = & \mathbb{E}\sum_{j=0}^{L_{n}}q_{j}^{-1}=\mathbb{E}t_{B}+q_{0}^{-1}\end{eqnarray*}
 and taking second derivative at $a=1_+$, we have \begin{eqnarray*}
n(n-1)+2nq_{0}^{-1} & = &
\mathbb{E}(\sum_{j=0}^{L_{n}}q_{j}^{-1})^{2}-\mathbb{E}\sum_{j=0}^{L_{n}}q_{j}^{-2}\\
 & = & \mathbb{E}(t_{B}+q_{0}^{-1})^{2}-\mathbb{E}(q_{0}^{-2}+t_{B}+C^{-1}t_{B}^{2}).\end{eqnarray*}
 Therefore, $\mathbb{E}t_{B}=n$ and $\mathbb{E}t_{B}^{2}=n^{2}C/(C-1)$.
Thus \begin{eqnarray*}
var(t_{B}) & = & \frac{n^2}{C-1}.\end{eqnarray*}
\end{proof}

{\bf Remark.} This elegant martingale argument already appeared in
\cite{rosenkrantz.1987} but under a different and simpler setting, and we
rediscovered it.

In implementation, we use the truncated version of $B$,
i.e. \eqref{eq:truncB}, which is equivalent to truncating the
theoretical estimate by $N$ if it is greater than $N$. Since by
assumption the true
cardinalities are no more than $N$,  this truncation removes one-sided
bias and thus reduces the theoretical RRMSE as shown in the above
theorem. Our simulation below shows that this truncation effect is
practically ignorable.

\section{Simulation studies and comparison}

\label{sec:simu}

In this section, we first present empirical studies that justify the
theoretical analysis of S-bitmap.
Then we compare S-bitmap with state-of-the-art algorithms in the literature
in terms of
memory efficiency and the scale invariance property.

\begin{figure*}[tt!]
\centering
\includegraphics[width=2.5in,height=2.5in,angle=-90]{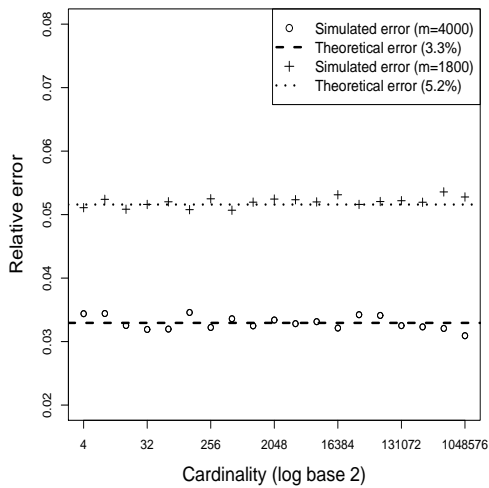}
\caption{Empirical and theoretical estimation errors of S-bitmap with $m=4,000$
bits and $m=1,800$ bits of memory for estimating cardinalities $1\leq
n\leq 2^{20}$.}\label{fig:fixNC}
\end{figure*}

\subsection{Simulation validation of S-bitmap's theoretical performance}\label{subsec:simustable}

In the above, our theoretical analysis shows that without truncation
by $N$, the
S-bitmap has a scale-invariant relative error
$\epsilon=(C-1)^{-1/2}$ for $n$ in a wide range $[1,N]$, where $C$ satisfies
Equation \eqref{eq:m} given bitmap size
$m$. We study the S-bitmap estimates based on \eqref{eq:truncB} with two sets
of simulations, both with $N=2^{20}$ (about one million), and then
compare empirical errors with the theoretical results. In
the first set, we fix $m=4,000$, which gives
$C=915.6$ and $\epsilon=3.3\%$, and in the second set, we fix
$m=1,800$, which gives $C=373.7$ and $\epsilon=5.2\%$.
We design the sequential
sampling rates according  to Section \ref{subsec:dimrule}.
For $1\leq n\leq N$,
we simulate $n$ distinct items and obtain S-bitmap estimate.
For each $n$ (power of 2), we replicate the simulation 1000 times and
obtain the empirical RRMSE.
These empirical errors are
compared with the theoretical errors in Figure
\ref{fig:fixNC}.
The results show that for both sets, the empirical errors and theoretical
errors match extremely well and the truncation effect is hardly visible.

\subsection{Comparison with state-of-the-art algorithms}
In this subsection, we demonstrate
that S-bitmap is more efficient in terms of memory and accuracy, and
more reliable than state-of-the-art algorithms such as mr-bitmap,
LogLog and Hyper-LogLog for many practical settings.

\begin{table}
\centering
{
\begin{tabular}[htbp]{|l|c|c|c|c|c|c|c|}
\hline
$N$            & \multicolumn{2}{|c|}{$\epsilon=1\%$ }   &
\multicolumn{2}{|c|}{$\epsilon=3\%$ } & \multicolumn{2}{|c|}{$\epsilon=9\%$}\\
\hline
                 & HLLog & S-bitmap & HLLog & S-bitmap & HLLog & S-bitmap\\
\hline

$10^3$ & 432.6 & 59.1 & 48.1 & 11.3 & 5.3 & 2.4\\
\hline
$10^4$ & 432.6 & 104.9 & 48.1 & 21.9 & 5.3 & 3.8\\
\hline
$10^5$ & 540.8 & 202.2 & 60.1 & 34.5 & 6.7 & 5.2\\
\hline
$10^6$ & 540.8 & 315.2 & 60.1 & 47.2 & 6.7 & 6.6\\
\hline
$10^7$ & 540.8 & 430.1 & 60.1 & 60 & 6.7 & 8.1\\
\hline
\end{tabular}
}
\caption{Memory cost (with unit 100 bits) of
  Hyper-LogLog and S-bitmap with given $N,\epsilon$.}
\label{tb:sloglog}
\end{table}

\begin{figure}[tt!]
\centering
\begin{tabular}{c}
\includegraphics[width=2.5in,height=3in,angle=-90]{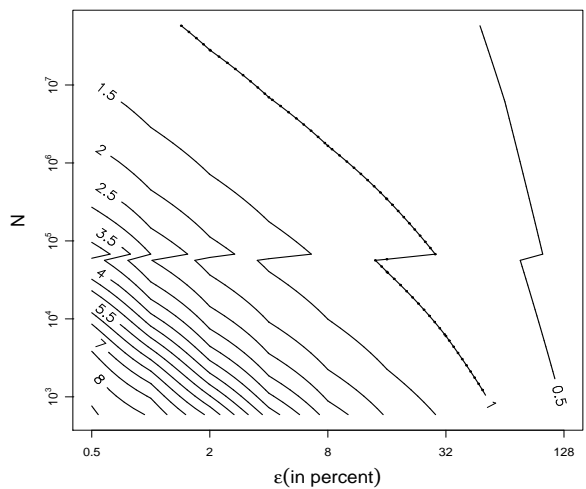}\\
\end{tabular}
\caption{ Contour plot of the ratios of the memory cost of
  Hyper-LogLog to that of S-bitmap with the same $(N,\epsilon)$: the
  contour line with small circles and label '1' represents the contour with ratio values equal to 1.}
\label{fig:sbitmap-hypera}
\end{figure}

\vspace{.15in}
\noindent{\bf Memory efficiency} Hereafter, the memory cost of a
distinct counting algorithm stands for the size
of the summary statistics (in bits) and does not count for hash
functions (whose seeds require some small memory space), and we note that the
algorithms to be compared here all require at least one universal hash
function. From \eqref{eq:m},
the memory cost for S-bitmap
is approximately linear in $\log (2N/C)$.
By the theory developed in \cite{durand.flajolet.2003} and \cite{flajolet.et.al.07},
the space requirements for LogLog counting and Hyper-LogLog are
approximately
$1.30^{2}\times \alpha\epsilon^{-2}$ and
$1.04^{2}\times \alpha\epsilon^{-2}$ in order to achieve
RRMSE $\epsilon=(C-1)^{-1/2}$, where
\begin{eqnarray*}
\alpha & = & 5, \text{ if }2^{16}\leq N <2^{32},\\
       & = & 4,  \text{ if }2^{8}\leq N <2^{16}.
\end{eqnarray*}
Here $\alpha=k+1$ if $2^{2^k}\leq N < 2^{2^{k+1}}$ for any positive integer $k$. So
LogLog requires
about 56\% more memory than Hyper-LogLog to achieve the same asymptotic
error. There is no analytic study
of the memory cost for mr-bitmap in the literature, thus
below we report a thorough
memory cost comparison only between S-bitmap and Hyper-LogLog.

Given $N$ and $\epsilon$, the theoretical memory costs for S-bitmap
and Hyper-LogLog can be calculated as above.
Figure
\ref{fig:sbitmap-hypera} shows the contour plot of the ratios
of
the memory requirement of Hyper-LogLog to that of S-bitmap, where the ratios
are shown as the labels of corresponding contour lines.
Here $\epsilon\times 100\%$ is
shown in the horizontal axis and $N$ is shown in the vertical axis,
both in the scale of
log base 2.  The contour line with small circles and label '1' shows the boundary where
Hyper-LogLog and S-bitmap require the same memory cost $m$. The lower left
side of this contour line is the region where Hyper-LogLog
requires more memory than S-bitmap, and the upper right side shows the
opposite.
Table \ref{tb:sloglog} lists the detailed memory cost for both
S-bitmap and Hyper-LogLog in a few cases where
$\epsilon$ takes values 1\%, 3\% and 9\%, and $N$
takes values from 1000 to $10^8$.
For example, for $N=10^6$ and $\epsilon\leq 3\%$, which is a suitable
setup for a
core network flow monitoring, Hyper-LogLog requires
at least 27\% more memory than S-bitmap.
As another example, for $N=10^4$ and $\epsilon\leq 3\%$, which is a reasonable
setup for household network monitoring, Hyper-LogLog requires at least 120\% more memory than S-bitmap.
In summary,
S-bitmap is uniformly more memory-efficient than Hyper-LogLog
when $N$ is medium or small and $\epsilon$ is small, though the
advantage of S-bitmap against Hyper-LogLog dissipates with $N\geq
10^7$ and large $\epsilon$.

\begin{figure}[tt!]
\centering
\includegraphics[width=4in,height=5in,angle=-90]{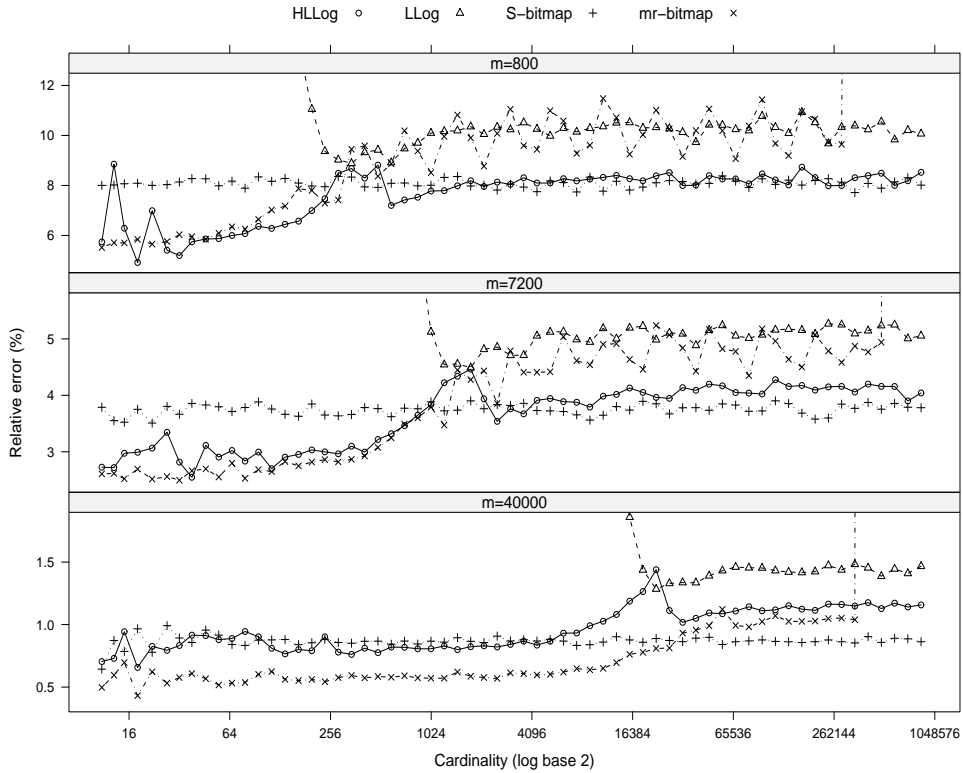}
\caption{Comparison among mr-bitmap,
  LogLog, Hyper-LogLog and S-bitmap for estimating cardinalities from
  10 to $10^6$ with  $m=40,000$, $m=3,200$ and $m=800$ respectively.}
\label{fig:simucomp}
\end{figure}

\vspace{.15in}
\noindent{\bf Scale-invariance property}
In many applications, the cardinalities of interest are in the scale
of a million or less. Therefore we report simulation studies with
$N=2^{20}$.
In the first experiment, $m=40,000$ bits of
memory is used for all four algorithms. The design of mr-bitmap is optimized
according to \cite{bitmap:2006}.
Let the true cardinality $n$ vary from 10 to $10^6$ and  the
algorithms are run to obtain corresponding estimates $\hat{n}$ and estimation
errors $n^{-1}\hat{n}-1$.
Empirical RRMSE is computed based on 1000
replicates of this procedure.
In the second and third experiments, the setting is similar except
that $m=3,200$ and $m=800$ are used, respectively.
The performance comparison is reported in
Figure~\ref{fig:simucomp}. The results show that in the first
experiment, mr-bitmap has small errors than LogLog and HyperLogLog,
but S-bitmap has smaller errors than all competitors for cardinalities
greater than 40,000; In
the second experiment,  Hyper-LogLog performs better than
mr-bitmap, but S-bitmap performs better than all competitors
for cardinalities greater than
1,000; And in the third experiment, with higher errors, S-bitmap still
performs slightly
better than Hyper-LogLog for cardinalities greater than 1,000, and
both are better than mr-bitmap and LogLog. Obviously, the scale invariance
property is validated for S-bitmap consistently, while it is not the case for
the competitors.  We note that mr-bitmap performs badly at the boundary, which are
not plotted in the figures as they are out of range.

\vspace{.15in}

\vspace{.15in}
\noindent{\bf Other performance measures}
Besides RRMSE, which is the $L_2$ metric, we have also evaluated the performance based on other metrics such as $\mathbb{E}|n^{-1}\hat{n}-1|$, namely the $L_1$ metric, and the quantile of $|n^{-1}\hat{n}-1|$. As examples, Table \ref{tb:N10K}  and Table \ref{tb:N1M} report the comparison of three error metrics ($L_1$, $L_2$ and 99\% quantile) for the cases with ($N=10^4$, $m=2700$) and ($N=10^6$, $m=6720$), which represent two settings of different scales. In both settings, mr-bitmap works very well for small cardinalities and worse as cardinalities get large, with strong boundary effect. Hyper-LogLog has a similar behavior, but is much more reliable. Interestingly, empirical results suggest that the scale-invariance property holds for S-bitmap not only  with RRMSE, but approximately with the metrics of $L_1$ and the 99\% quantile. For large cardinalities relative to $N$, the errors of Hyper-LogLog are all higher than that of S-bitmap in both settings.

\begin{table} \centering
\begin{tabular}{|c|ccc|ccc|ccc|ccc|}
\hline
& \multicolumn{3}{|c|}{$L_1$} & \multicolumn{3}{|c|}{$L_2$ (RRMSE)} & \multicolumn{3}{|c|}{99\% quantile} \\ 
$n$ & S & mr & H & S & mr & H &S & mr & H \\ \hline

10 & 1.3 & 0.6 & 0.8 & 2.6 & 1.6 & 3 & 10 & 10 & 10 \\ \hline 
100 & 2.1 & 1.4 & 2.5 & 2.6 & 1.7 & 3.2 & 6 & 4 & 8 \\ \hline 
1000 & 2.1 & 1.6 & 3.5 & 2.6 & 2 & 4.4 & 6.7 & 5 & 11.4 \\ \hline 
5000 & 2.1 & 2.3 & 3.4 & 2.6 & 3.4 & 4.2 & 6.6 & 7.5 & 11.3 \\ \hline 
7500 & 2.1 & 100.7 & 3.5 & 2.6 & 100.9 & 4.3 & 6.9 & 119 & 11.2 \\ \hline 
10000 & 2.1 & 101.9 & 3.5 & 2.6 & 102.4 & 4.4 & 6.6 & 131.1 & 11.5 \\ \hline 
\end{tabular}
\caption{Comparison of  $L_1$, $L_2$ metrics and 99\%-quantiles (times 100) among mr-bitmap (mr), Hyper-LogLog (H) and S-bitmap (S) for $N=10^4$ and $m=2700$.}
\label{tb:N10K}
\end{table}

\begin{table} 
\centering
\begin{tabular}{|c|ccc|ccc|ccc|ccc|}
\hline

& \multicolumn{3}{|c|}{$L_1$} & \multicolumn{3}{|c|}{$L_2$ (RRMSE)} & \multicolumn{3}{|c|}{99\% quantile} \\ 
$n$ & S & mr & H & S & mr & H &S & mr & H \\ \hline

10 & 1.1 & 0.5 & 0.4 & 2.4 & 1.3 & 1.9 & 10 & 10 & 10 \\ \hline 
100 & 1.8 & 1.4 & 1.6 & 2.3 & 1.7 & 2 & 6 & 4 & 5 \\ \hline 
1000 & 1.9 & 1.5 & 1.8 & 2.4 & 1.9 & 2.2 & 6.2 & 5 & 5.5 \\ \hline 
10000 & 2 & 2.5 & 2.1 & 2.5 & 3.1 & 2.7 & 6.8 & 7.9 & 7 \\ \hline 
1e+05 & 1.9 & 2.6 & 2.3 & 2.4 & 3.3 & 2.9 & 6.5 & 7.9 & 7.6 \\ \hline 
5e+05 & 1.9 & 2.6 & 2.3 & 2.4 & 3.3 & 2.8 & 6.2 & 8.6 & 7.3 \\ \hline 
750000 & 2 & 22.9 & 2.2 & 2.5 & 48.2 & 2.8 & 6.1 & 116.9 & 7 \\ \hline 
1e+06 & 1.9 & 100.5 & 2.2 & 2.4 & 100.8 & 2.8 & 6.2 & 120.3 & 7.4 \\ \hline 

\end{tabular}
\caption{Comparison of $L_1$, $L_2$ metrics and 99\%-quantiles (times 100) among mr-bitmap (mr), Hyper-LogLog (H) and S-bitmap (S) for $N=10^6$ and $m=6720$.}
\label{tb:N1M}

\end{table}

\section{Experimental evaluation}

\label{sec:experiment}

We now evaluate the S-bitmap algorithm on a few real network data
and also compare it with the three competitors as above.

\begin{figure*}[tt!]
\centering
\begin{tabular}{cc}
\parbox{2.5in}{\centering
\includegraphics[width=2.5in,height=2.5in,angle=-90]{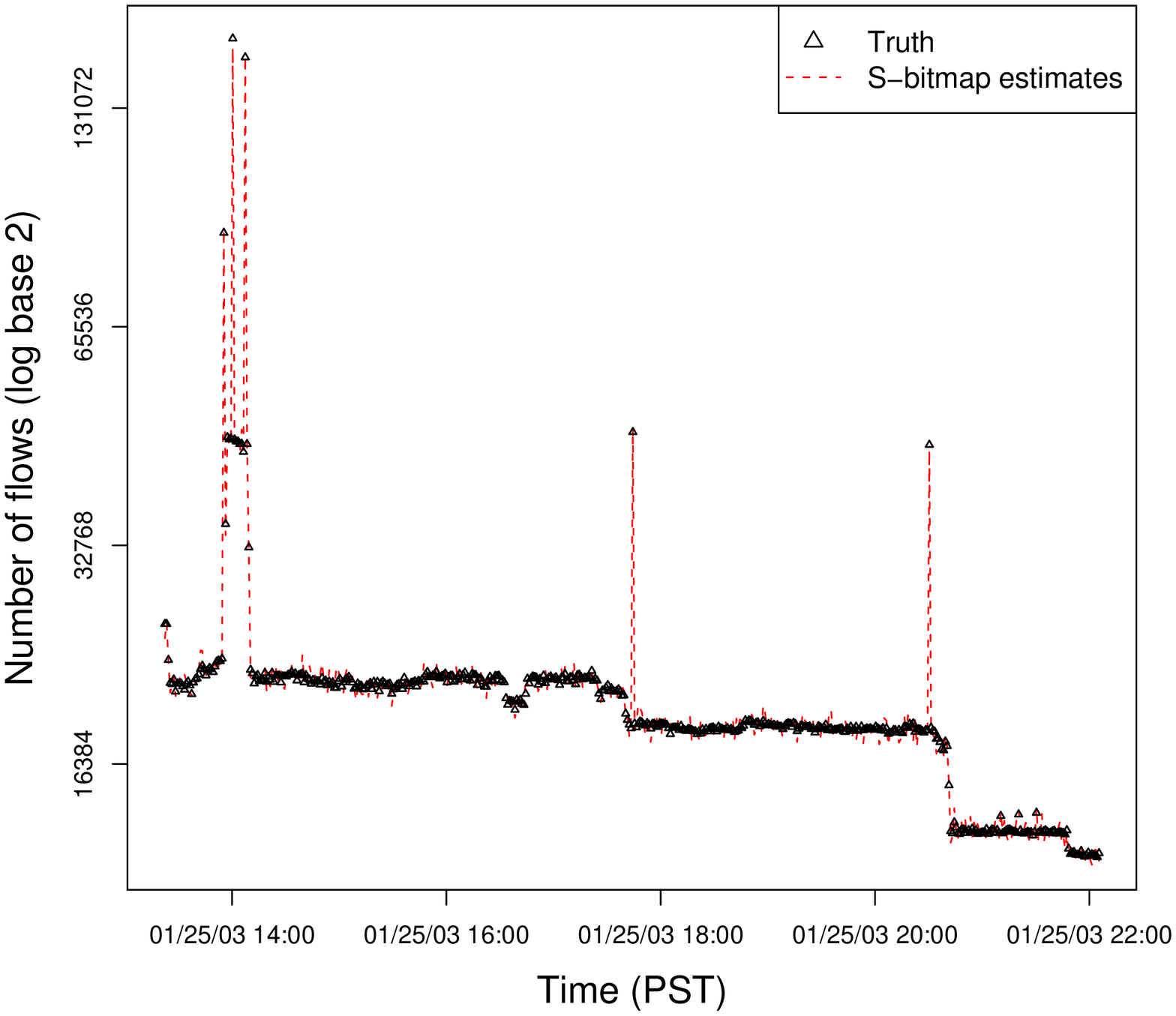}}
&
\parbox{2.5in}{\centering
\includegraphics[width=2.5in,height=2.5in,angle=-90]{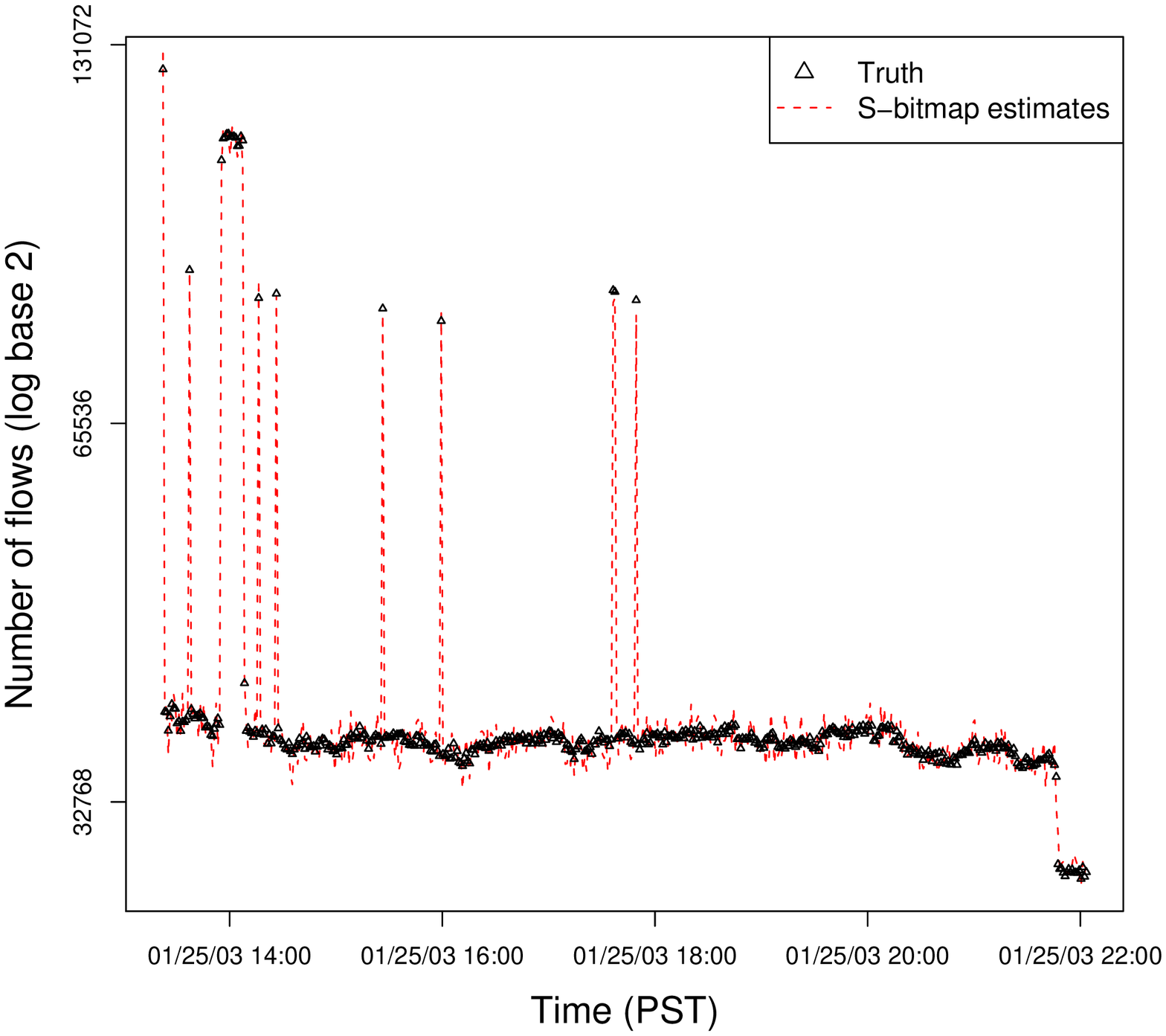}}
\\
\mbox{\small (a) Link 1} &
\mbox{\small (b) Link 0}
\end{tabular}
\caption{Time series of true flow counts (in triangle) and
   S-bitmap estimates (in dotted line) per minute on both links during
   slammer outbreak: link 1
   (a) and link 0 (b).}
\label{fig:timeseries10}
\end{figure*}

\begin{figure*}[tt!]
\centering
\begin{tabular}{cc}
\parbox{2.5in}{\centering
\includegraphics[width=2.5in,height=2.5in,angle=-90]{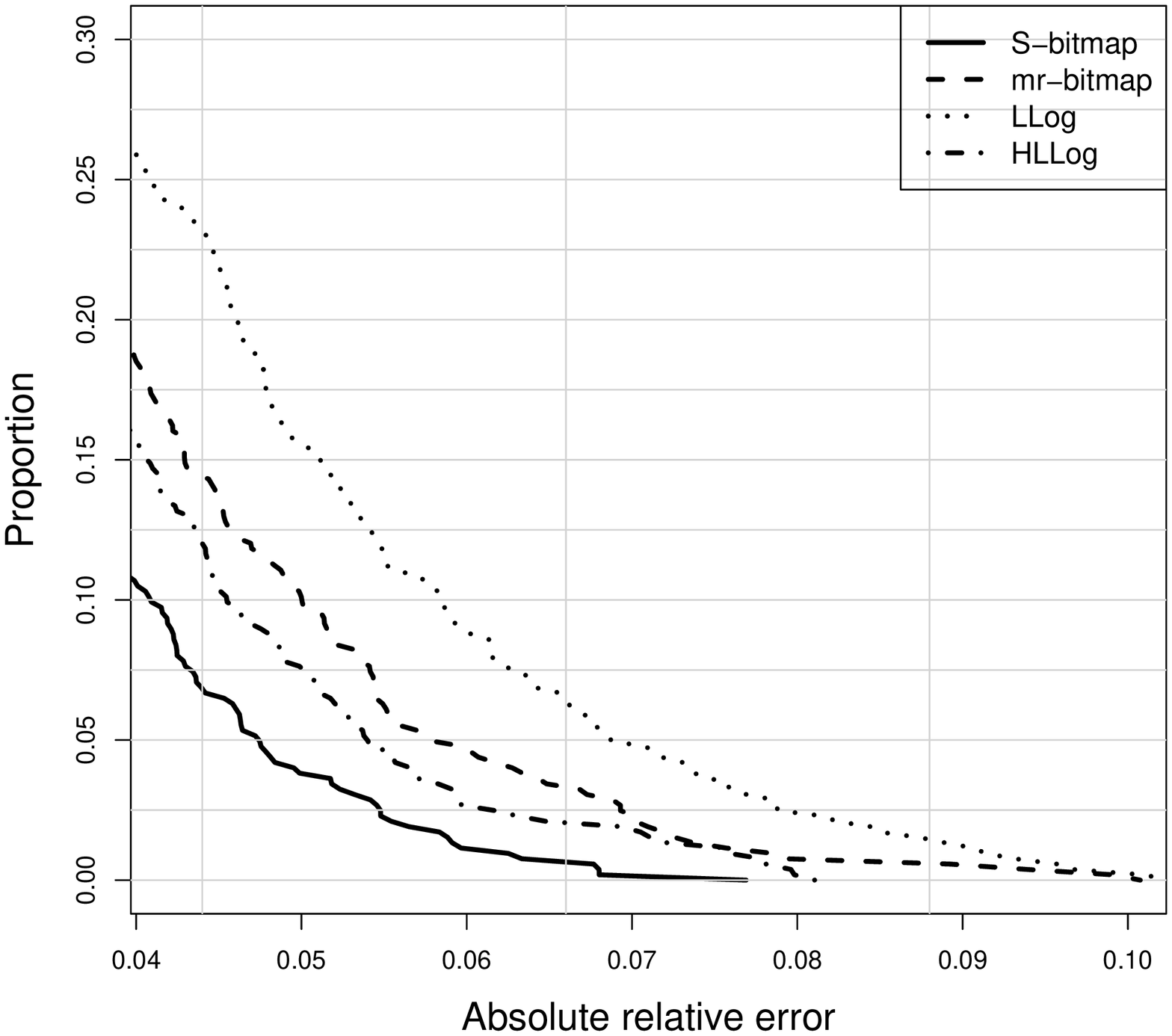}}
&
\parbox{2.5in}{\centering
\includegraphics[width=2.5in,height=2.5in,angle=-90]{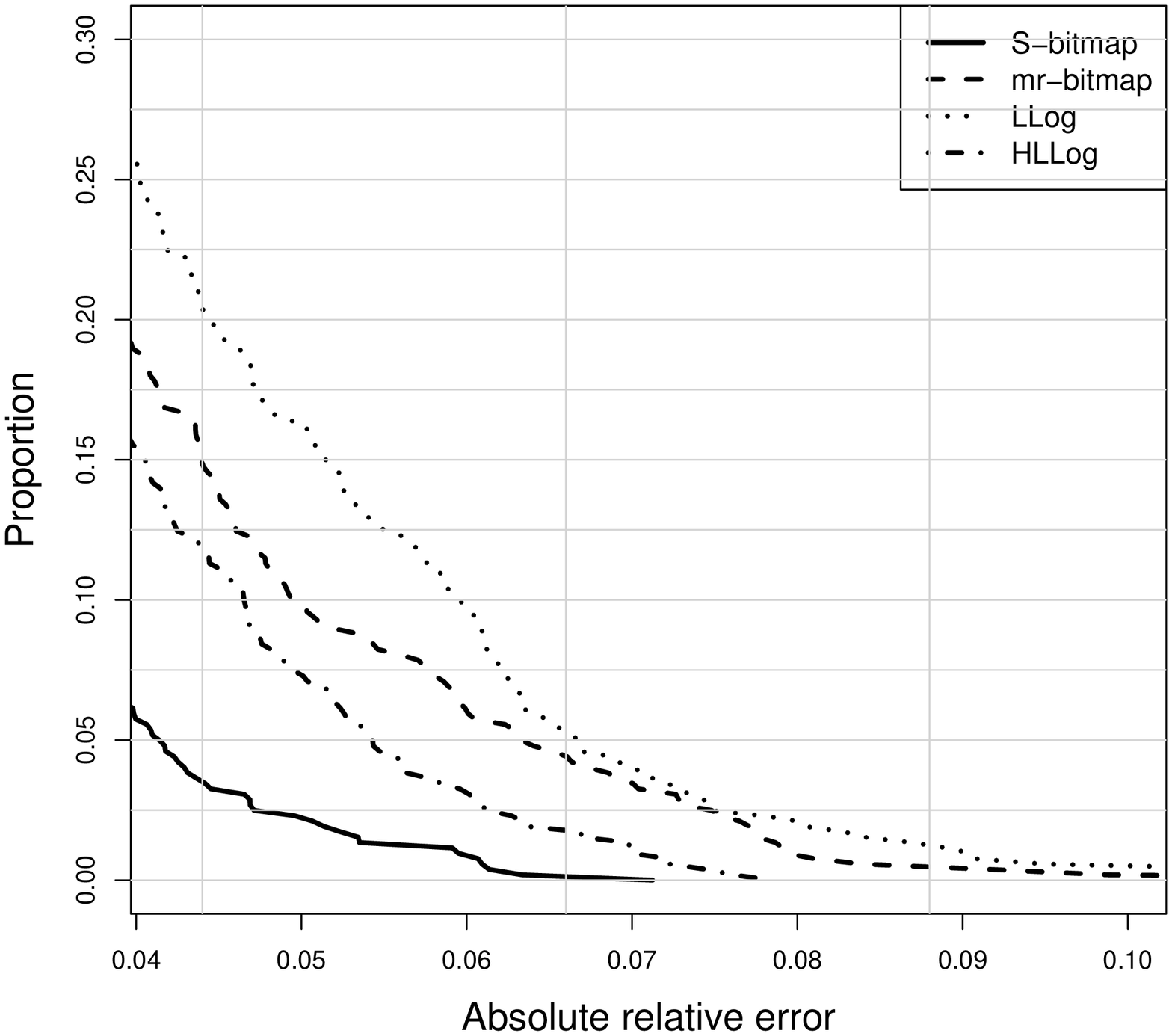}}
\\
\mbox{\small (a) Link 1} &
\mbox{\small (b) Link 0}
\end{tabular}
\caption{Proportions of estimates (y-axis) that have RRMSE more than a
  threshold (x-axis) based on
  S-bitmap,
  mr-bitmap, LogLog and Hyper-LogLog, respectively on the two links
  during slammer outbreak: link 1 (a) and link 0 (b), where the three
  vertical lines show 2, 3 and 4 times expected standard deviation for
  S-bitmap separately.}
\label{fig:wormquantile}
\end{figure*}

\subsection{Worm traffic monitoring}
We first evaluate the algorithms on worm traffic data, using two
9-hours traffic traces (www.rbeverly.net/research/slammer). The traces
were collected
by MIT Laboratory for Computer Science from a peering exchange point
(two independent
links, namely link 0 and link 1) on Jan 25th 2003, during the period
of ``Slammer`` worm outbreak.
We report the results of estimating flow
counts for each link. We take $N=10^6$, which is sufficient for most
university traffic in normal scenarios. Since in
practice routers may not allocate much
resource for flow counting, we use $m=8000$ bits. According
to \eqref{eq:m}, we obtain
$C=2026.55$ for designing the sampling rates for S-bitmap, which
corresponds to an expected standard deviation of
$\epsilon = 2.2\%$ for S-bitmap. The
same memory is used for other algorithms. The two
panels of Figure \ref{fig:timeseries10} show the
time series of flow counts every minute interval in triangles on link 1
and link 0 respectively, and the corresponding S-bitmap estimates in
dashed lines. Occasionally the flows become very bursty (an order
of difference), probably due to a few heavy worm scanners, while most
times the time series are pretty stable.
The estimation errors of
the S-bitmap estimates are almost invisible despite the non-stationary
and bursty points. 

The performance comparison between S-bitmap and alternative methods is
reported in Figure \ref{fig:wormquantile}
(left for Link 1 and right for Link 0),
where y-axis is the proportion of
estimates that have absolute relative estimation errors more than a given
threshold in the x-axis.
The three thin vertical lines show the 2, 3 and 4 times expected
standard deviation for S-bitmap, respectively. For example,
the proportion of S-bitmap
estimates whose absolute relative errors are more than 3
times the expected standard deviation is almost 0 on both links, while
for the competitors, the proportions are at least 1.5\%
given the same threshold.
The results show that
S-bitmap is most resistant to large errors among all four
algorithms for both Link 1 and Link 0.

\begin{figure*}[ht]
\begin{center}
\includegraphics[width=2.5in,height=2.5in,angle=-90]{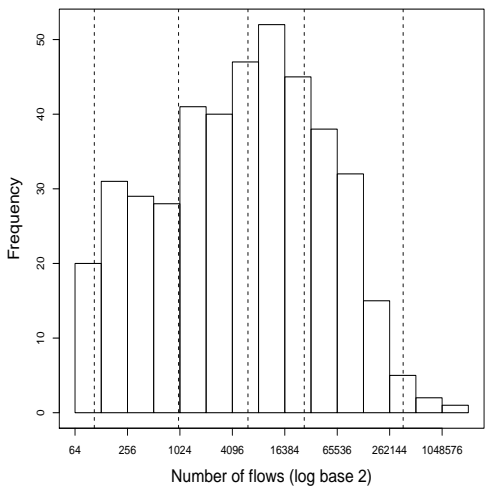}
\caption{Histogram of five-minute flow counts on backbone links (log
  base 2).
}\label{fig:GChist}
\end{center}
\end{figure*}

\begin{figure*}[ht]
\begin{center}
\includegraphics[width=2.5in,height=2.5in,angle=-90]{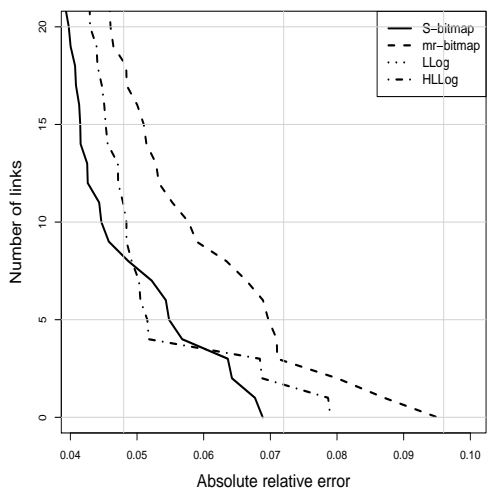}
\caption{Proportions of estimates (y-axis) that have RRMSE more than a
  threshold (x-axis) based on
  S-bitmap,
  mr-bitmap, LogLog and Hyper-LogLog, respectively, where the three
  vertical times show 2, 3 and 4 times expected standard deviation for
  S-bitmap, separately.
}\label{fig:GCcomp}
\end{center}
\end{figure*}

\subsection{Flow traffic on backbone network links}

Now we apply the algorithms for counting network link flows in a core network.
The real data
was obtained from a Tier-1 US service
provider for 600 backbone links in the core network, which includes time series
of traffic volume in flow counts on MPLS (Multi Protocol Label Switching)
paths in every five minutes.
The traffic scales vary dramatically from link to link as well as from
time to time.
Since the original
traces are not available, we use simulated data for each link
to compute S-bitmap and then obtain estimates. We set $N=1.5\times
10^6$ and use $m=7,200$ bits of memory to configure all algorithms as
above, which corresponds to an expected standard deviation of 2.4\%
for S-bitmap. The simulation uses a snapshot
of a five minute interval flow counts, whose histogram in log base
2 is presented in Figure \ref{fig:GChist}.
The vertical lines show that the .1\%, 25\%, 50\%, 75\%, and 99\%
quantiles are 18, 196, 2817, 19401 and 361485 respectively, where
about 10\% of the
links with no flows or flow counts less than 10 are not considered.
The performance comparison between S-bitmap and alternative methods is
reported in Figure \ref{fig:GCcomp}
similar to Figure \ref{fig:wormquantile}.
The results show that
both S-bitmap and Hyper-LogLog give very accurate estimates with
relative estimation errors bounded by 8\%, while mr-bitmap has
worse performance and LogLog is the worst (off the range). Overall,
S-bitmap is most resistant to large errors among all four
algorithms. For
example, the absolute relative errors based on
S-bitmap are within 3 times the standard deviation for all links,
while there is one link whose absolute relative
error is beyond this threshold  for Hyper-LogLog, and two such links
for mr-bitmap.

\section{Conclusion}

Distinct counting is a fundamental problem in the database literature
and has found important applications in many areas, especially in
modern computer networks. In this paper, we have proposed a
novel statistical solution
(S-bitmap), which is scale-invariant in the sense that its relative
root mean square error is independent of the unknown cardinalities in a
wide range.
To achieve the same accuracy, with similar computational cost,
S-bitmap consumes significantly less memory than
state-of-the-art methods
such as multiresolution bitmap, LogLog counting and  Hyper-LogLog for
common practice scales.  

\section*{Appendix}

\subsection{Proof of Lemma \ref{lem:geomlem}}

By the definition
of $\{ T_{k}:1\leq k\leq m\}$, we have \begin{eqnarray*}
 &  & \mathbb{P}(T_{k}-T_{k-1}=t)\\
 & = & \sum_{s=k-1}^{\infty}\mathbb{P}(T_{k-1}=s,T_{k}=t+s)\\
 & = & \sum_{s=k-1}^{\infty}\mathbb{P}(I_{s}=1,I_{t+s}=1,L_{s}=k-1,L_{t+s}=k).\end{eqnarray*}
 Since $L_{s}\leq L_{s+1}\leq\cdots\leq L_{s+t}$, by the Markov chain
property of $\{ L_{t}:t=1,\cdots,\}$, we have for $k\geq1$ and $s\geq k-1$,
\begin{eqnarray*}
 &  & \mathbb{P}(I_{s}=1,I_{t+s}=1,L_{s}=k-1,L_{t+s}=k)\\
 & = & \mathbb{P}(L_{s}=k-1,I_{s}=1)\mathbb{P}(L_{t+s}=k|L_{t+s-1}=k-1)\\
 &  & \times\prod_{j=s+1}^{s+t-1}\mathbb{P}(L_{j}=k-1|L_{j-1}=k-1)\\
 & = & \mathbb{P}(T_{k-1}=s)q_{k}\prod_{j=s+1}^{s+t-1}(1-q_{k})\\
 & = & \mathbb{P}(T_{k-1}=s)q_{k}(1-q_{k})^{t-1}.\end{eqnarray*}
 Notice that $\sum_{s=k-1}^{\infty}\mathbb{P}(T_{k-1}=s)=\mathbb{P}(T_{k-1}\geq k-1)$
is probability that the $(k-1)$-th filled bucket happens when or
after the $(k-1)$-th distinct item arrives, which is $100\%$ since
each distinct item can fill in at most one empty. Therefore \begin{eqnarray*}
\mathbb{P}(T_{k}-T_{k-1}=t) & = & q_{k}(1-q_{k})^{t-1}.\end{eqnarray*}
 That is, $T_{k}-T_{k-1}$ follows a geometric distribution. The independence
of $\{ T_{k}-T_{k-1}:1\leq k\leq m\}$ can be proved similarly using
the Markov property of $\{ L_{t}:t=1,2,\cdots\}$, which we refer
to Chapter 3 of \cite{durrett.1996}. This completes the proof of
Lemma \ref{lem:geomlem}.

\bibliographystyle{natbib}
\bibliography{abit}

\end{document}